\documentclass{elsart}

 \usepackage{graphics}
 \usepackage{graphicx}
 \usepackage{epsfig}

\usepackage{amssymb}

\begin{document}

\begin{frontmatter}


 %\title{}
%\thanks[label1]{1}
 %\author{}

% \address{}

\title{Special Attention Network}


\author{J. O. Indekeu}
%\corauthref{cor1}
\ead{joseph.indekeu@fys.kuleuven.ac.be}
%\corauth[cor1]{Corresponding author} and
%\author{}
%\ead{claudiu.giuraniuc@fys.kuleuven.ac.be}
\address{Laboratorium voor Vaste-Stoffysica en
Magnetisme,\\Celestijnenlaan 200 D, Katholieke Universiteit
Leuven,\\ B-3001 Leuven, Belgium}

\begin{abstract}
In this Note a social network model for opinion formation is
proposed in which a person connected to $q$ partners pays an {\em
attention} $1/q$ to each partner. The mutual attention between two
connected persons $i$ and $j$ is taken equal to the geometric mean
$1/\sqrt{q_iq_j}$. Opinion is represented as usual by an Ising
spin $s=\pm 1$ and mutual attention is given through a two-spin
coupling $J_{ij} = J Q/\sqrt{q_iq_j}$,  $Q$ being the average
connectivity in the network. Connectivity diminishes attention and
only persons with low connectivity can pay special attention to
each other leading to a durable common (or opposing) opinion. The
model is solved in ``mean-field" approximation and a critical
``temperature" $T_c$ proportional to $JQ$ is found, which is
independent of the number of persons $N$, for large $N$.

\end{abstract}

\end{frontmatter}

% main text
%\section{\large{Introduction}}

Recently Aleksiejuk {\em et al.}\cite{Alek} proposed a social
network model for opinion formation by introducing
ferromagnetically coupled Ising spins on a Barab\'asi-Albert
network \cite{Bar}. The strength of the couplings between linked
nodes, $J$, was taken to be uniform, independent of the number of
connections to a node. Simulations of this model indicated the
existence of a critical ``temperature" below which opinion
formation is possible (two-phase coexistence) and above which
common opinion is unstable (disordered phase). A peculiar feature
of this model is that simulations indicate that the critical
temperature depends on the number of nodes, or system size, $N$ in
the manner $T_c \propto \log(N)$, so that in the thermodynamic
limit an initially imposed common opinion always persists (in the
absence of an external opposing ``field"), no matter how weak the
uniform coupling $J$ between each pair of connected persons, or
``partners" \cite{Alek}. Incidentally, this peculiar divergence of
$T_c$ with $N$ is missed in the simplest mean-field approximation
to the model \cite{Alek}, but is captured correctly in an improved
mean-field approach \cite{Bian,Leon,Igloi}, and agrees with exact
results for uncorrelated networks \cite{Doro}.

In the model of Aleksiejuk {\em et al.} the assumption of uniform
coupling lends a lot of influence to nodes with many connections.
Indeed, overturning the spin of a small set of most heavily
connected nodes suffices for reversing the opinion of the entire
network. It is, however, questionable whether a node, e.g., a
person, with many partners (i.e., nodes to which he/she is linked)
is able to influence all these partners as strongly as a person
with only a few partners would be able to do. An intensive
person-to-person discussion presumably creates a stronger tendency
to form a durable common (or, for antiferromagnetic couplings,
opposing) opinion than a one-to-many communication. Therefore we
propose to attribute an ``attention" to each person, inversely
proportional to the number of partners. In network terms, a node
with connectivity $q_i$ is capable of maintaining an attention
\begin{equation}
\alpha_i
 = 1/q_i
\end{equation}
towards each of its partners, the total attention per node being
normalized to 1. Thus we assume that each person pays the same
total attention, $q_i \alpha_i = 1$, to the exterior. If the
average connectivity is denoted by $Q$, $q_i < Q$ signifies
``special attention" and $q_i
> Q$ implies ``little attention". The model further assumes that
in order to establish a strong mutual influence of opinion, the
average of the attentions of two connected persons must be
sufficiently big. Thus the average of their connectivities, $q_i$
and $q_j$, must be small enough. In particular, if both partners
pay special attention to each other they are more likely to
maintain durable agreement (or disagreement). In contrast, if they
pay little attention to each other, opinion formation is
difficult. An interesting mixed case is that of a person $i$
paying special attention ($q_i < Q$) to TV or other mass media $j$
($q_j >> Q$). Although the person may quickly form an opinion
dictated by the mass medium, this opinion is not strengthened by
loyalty or peer-pressure considerations and may quickly ``flip".
For example, the person soon realizes that while he/she may be
especially devoted to TV (high $\alpha_i$), TV cares little or
nothing about him/her individually (low $\alpha_j$), which weakens
considerably the persuasive power of the medium.

Along this line of reasoning we advocate that reciprocity of
attention is important for durable opinion formation and an
interaction is proposed which is proportional to the mean - for
calculational simplicity the geometric mean - of the attentions of
the connected persons,
\begin{equation}
J_{ij} = J Q \sqrt{\alpha_i \alpha_j}
\end{equation}
Note that the average coupling equals $J$, within a
mean-connectivity approximation ($q_i = Q$), which in the present
social context is called a mean-attention approximation. Taking
the geometric mean leads to separable couplings, which usually
facilitates calculations drastically (cf. separable spin-glass
models \cite{Gren}). Also note that, in this symmetric model,
there is no directionality in the couplings, $J_{ij}= J_{ji}$. At
this stage no distinction is made between attention as ``speaker"
or as ``listener".

For a start, it is straightforward to apply a double mean-field
approximation as follows. For a given network realization
(quenched randomness) the exact self-consistent ``equation of
state" reads
\begin{equation}
 <s_i> = <\tanh ( \sum_{j=1}^{q_i} \frac{J_{ij}}{k_BT} s_j)> ,
 \end{equation}
  where the bracket denotes thermal average and $k_B$ is the Boltzmann
constant. Now we perform the quenched random average over all
networks simply in the mean-attention approximation $\alpha_i
\approx 1/Q$, and we
 also invoke the mean-opinion approximation $s_i \rightarrow <s_i> \approx S$, with
$-1<S<1$, where $S$ is the average opinion. This gives
\begin{equation}
S =  \tanh ( S JQ/k_BT ),
\end{equation}
which leads to the familiar critical ``temperature" $T_c = JQ/k_B$
proportional to the average connectivity $Q$.

In a more refined step we apply the improved mean-field approach
of Bianconi \cite{Bian} and Leone {\em et al.}\cite{Leon}. In this
calculation an equation of state for the mean local opinion
$<s_i>$ is found, of the form
\begin{equation}
<s_i> = \tanh (\sum_{j=1}^N \frac{[J_{ij}]}{kT}<s_j>),
\end{equation}
where the square brackets denote the {\em quenched random average}
over network realizations. Note that now the sum runs over all
nodes. Specifically, for the present model, and assuming a
Barab\'asi-Albert network,
\begin{equation}
 [J_{ij}] = J_{ij}p_{ij} = \frac{J}{N}\sqrt{q_i q_j},
 \end{equation}
 where $p_{ij}$ is the probability that nodes $i$ and $j$ are
 linked.

 It is then natural to define the order parameter
 \begin{equation}
\hat S = \frac{1}{\sqrt{Q} N} \sum_{j=1}^N \sqrt{q_j} <s_j>
\end{equation}
Following Bianconi's continuum approximation \cite{Bian} the
critical temperature is then, for large $N$, obtained from the
linearized equation
\begin{equation}
\hat S = \frac{1}{N} \int_1^N dn' \frac{J}{k_BT} \hat S q(n'),
\end{equation}
with, for the Barab\'asi-Albert network in the large-$N$ limit
\cite{Bar},
\begin{equation}
q(n) \sim \frac{Q}{2} \sqrt{\frac{N}{n}},
\end{equation}
where $q(n)$ stands for the connectivity of node $n$. This also
leads to the same result $T_c = JQ/k_B$.

In view of the fact that none of the spins on this network exerts
an anomalously strong influence on other parts of the network, one
may expect that this model behaves normally, so that, when
fluctuations are taken into account, a finite critical temperature
results which is independent of $N$, in the large network limit.
Indeed, in the model of Aleksiejuk {\em et al.} a highly connected
spin $s_i$ exerts a local field of strength $J$ on a large set of
$q_i$ other spins. Consequently, for large $N$ such centers of
massive influence may well induce significant deviations from a
``well-behaved" thermodynamic limit, as is corroborated by the
simulations \cite{Alek}.

In sum, an opinion formation model has been proposed in which each
person can devote a fixed total amount of attention to others,
distributing this attention equally over all partners. The Special
Attention Network presented here attenuates the strong influence
exerted by highly connected nodes in networks with uniform
couplings $J$, by introducing a detailed local compensation of
high connectivity by weak interaction. Therefore, we conjecture
that the essential feature $T_c \propto JQ$ captured here in the
mean-field approximation(s) holds true also for the fluctuating
spin model on any quenched random Special Attention Network,
scale-free or not, with finite mean connectivity $Q < \infty$.
Careful Monte Carlo simulation and/or more sophisticated
analytical calculation will be needed to verify this.

%\begin{figure}[htbp]
%\centerline{\epsfxsize=15cm \epsfbox{}} \caption{}
%\end{figure}

{\bf Acknowledgments.}\\  This research is supported by the
Flemish Programme FWO-G.0222.02 ``Physical and interdisciplinary
applications of novel fractal structures". The author thanks
warmly Dietrich Stauffer, Carlo Vanderzande and Claudiu Giuraniuc
for encouraging remarks, informative discussions and pointing out
pertinent references. He is furthermore grateful to the Organizers
of the XVIIIth Max Born Symposium, L\c{a}dek Zdr\'oj, Poland, at
which this development was conceived.

\label{}

% \appendix

% \section{}
% \label{}

\end{document}